\documentclass[showpacs,twocolumn,superscriptaddress,aps,prl]{revtex4-1}

\usepackage{graphicx}
\usepackage{amsmath,amssymb,bm}
\usepackage{epstopdf}

\newcommand{\iChEM}{{\it i}{\rm ChEM}}

\newcommand{\B}{\text{\scriptsize res}}
\newcommand{\s}{\text{\scriptsize sys}}
\newcommand{\SB}{\text{\scriptsize sys-res}}
\newcommand{\T}{{\rm total}}
\newcommand{\dg}{\dagger}

\newcommand{\up}{\uparrow}
\newcommand{\down}{\downarrow}

\newcommand{\w}{\omega}
\newcommand{\M}{\text{\tiny M}}
\newcommand{\CT}{\text{\tiny CT}}
\newcommand{\DD}{\text{\tiny DD}}

\newcommand{\be}{\begin{equation}}
\newcommand{\ee}{\end{equation}}
\newcommand{\bea}{\begin{eqnarray}}
\newcommand{\eea}{\end{eqnarray}}
\newcommand{\bsube}{\begin{subequations}}
\newcommand{\esube}{\end{subequations}}

\newcommand{\Fig}[1]{Fig.\,\ref{#1}}

\begin{document}

\title{Many--body Tunneling and Nonequilibrium Dynamics of Doublons in Strongly Correlated Quantum Dots}

\author{WenJie Hou}
 \affiliation{Department of Physics, Renmin University of China, Beijing 100872, China}

\author{YuanDong Wang}
 \affiliation{Department of Physics, Renmin University of China, Beijing 100872, China}

\author{JianHua Wei}\email{wjh@ruc.edu.cn}
 \affiliation{Department of Physics, Renmin University of China, Beijing 100872, China}

\author{ZhenGang Zhu}\email{zgzhu@ucas.ac.cn}
 \affiliation{School of Electronic, Electrical and Communication Engineering, University of Chinese Academy of Sciences, Beijing 100049, China}

\author{YiJing Yan}\email{yanyj@ustc.edu.cn}
 \affiliation{Hefei National Laboratory for Physical Sciences at the Microscale and
 \iChEM\ (Collaborative Innovation Center of Chemistry for Energy Materials),
 University of Science and Technology of China, Hefei, Anhui 230026, China}

\date{\today}

\begin{abstract}

 Quantum tunneling dominates coherent transport at low temperatures in many systems of great interest. In this work we report a many--body tunneling (MBT), by nonperturbatively solving the Anderson multi-impurity model, and identify it a fundamental tunneling process on top of the well--acknowledged sequential tunneling and cotunneling. We show that the MBT involves the dynamics of doublons in strongly correlated systems. Proportional to the numbers of dynamical doublons, the MBT can dominate the off--resonant transport in the strongly correlated regime. A $T^{3/2}$--dependence of the MBT current on temperature is uncovered and can be identified as a fingerprint of the MBT in experiments. We  also prove that the MBT can support the coherent long--range tunneling of doublons, which is well consistent with recent experiments on ultracold atoms. As a fundamental physical process, the MBT is expected to play important roles in general quantum systems.

\end{abstract}

\maketitle

\begin{figure} 
\includegraphics[width=0.5\textwidth]{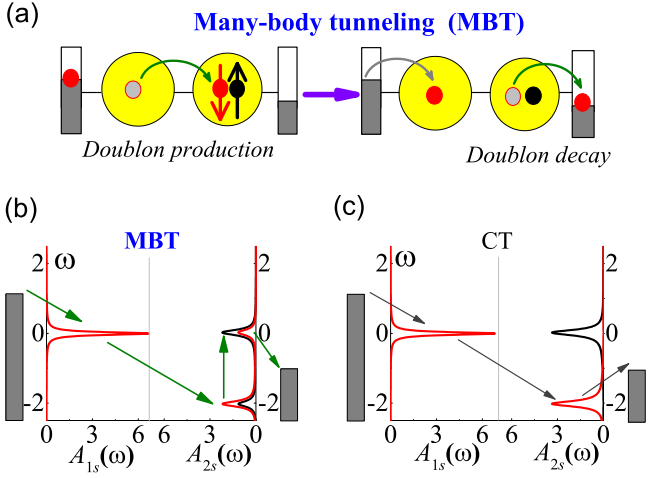}
\caption{(Color online).  (a) Schematic diagram of the many--body tunneling (MBT) in DQDs.
(b) and (c) show calculated nonequilibrium spectral functions $A_{1s}(\omega)$ and $A_{2s}(\omega)$ with $s=\down$ (red) and $s=\up$ (black), and illustrated processes of the MBT and CT respectively.
The accurate HEOM is performed in (b) and the HEOM truncated at tier level $L=1$ in (c) at $t=0.01$ meV, and $eV=U=2.0$ meV. $\epsilon_{2\down}=\epsilon_{2\up}=-2.0$ meV. Due to the lifted spin degeneracy of QD1, $A_{1\down(\uparrow)}(\w)$   peaks at  $\epsilon_{1\down}=0$ meV ($\epsilon_{1\up}=3.0$ meV (not shown)).  The dot-reservoir coupling strength and temperature are $\Gamma=0.1$ meV and $T=0.1$ meV, respectively.
}\label{fig1}
\end{figure}

{\it Introduction.}---Quantum tunneling is ubiquitous in quantum
systems \cite{Kouwenhoven943443,Vinay63486,Yukio953451,Jonathan963830,Hwang962041,Folling071029,Meinert141259}. Two basic
tunneling processes have been investigated: the first--order sequential tunneling (ST)
and the second--order cotunneling (CT) \cite{Gra92,Naz09,Averin902446}. Theoretically, both ST and CT can be well described by
a single--particle picture. On the other hand, quantum tunneling deeply involved many--body interactions (shortened with many-body tunneling (MBT)) inevitably exists and
dominates in many strongly correlated systems.
However, our understanding of the MBT is severely hindered by the difficult nonperturbative treatment of the many-body dynamics.  In this letter, we uncover a novel MBT which is involved with coherent production and decay of doublons (\Fig{fig1}(a)) and inherent with nonperturbative and nonequilibrium characters.

Doublon was initially proposed for the state of doubly-occupied site in Hubbard model \cite{Hub63283}, an excitation with respect to the half-filled Mott-insulator ground state. In strongly correlated systems, the doublon-holon (unoccupied sites) binding plays important role \cite{Kap82889}, which was proved to be closely related to the Mott transition and high-temperature superconductivity \cite{Pre15235155}.  Recently, the Hubbard models  have been successfully simulated in ultracold atoms.
In 2006, doublons with long lifetimes were observed in ultracold $^{87}{\rm Rb}$ atoms \cite{Win06853}. Since then,  there have been extensive studies on the properties of doublons in Bose- and Fermi-Hubbard models in ultracold atoms \cite{Robert08204,Strohmaier10080401,Greif11145302,Chudnovskiy12085302,Meinert13053003,Ole14193003,Xia15316,Covey1611279}.
Among those, the dynamics of doublons is of special interest, which can facilitate the understanding of the far-from-equilibrium dynamics of strongly correlated systems. The experiments of Fermi-Hubbard model (ultracold $^{40}{\rm K}$ atoms) found the decay rate of doublon scaling as $\tau^{-1}\propto \exp(-U/t)$, where $U$ is the on-site electron-electron ($e-e$) interaction and $t$ is the tunneling coupling between nearest-neighbor sites \cite{Strohmaier10080401}.  Similar relation was theoretically verified in Bose-Hubbard model \cite{Chudnovskiy12085302}.  The exponential dependence clearly originates from the many-body
character of the dynamics of doublons, which is far beyond the single-particle picture in the strongly correlated regime ($U/t\gg 1$).

Although the doublon have been studied in ultracold atom systems, the connection to the MBT and its many-body nature are not fully appreciated. We demonstrate that the doublon dynamics can be described by two-particle density matrix and is a kind of MBT naturally originating from strong Coulomb interaction. We select multi-impurity Anderson model \cite{Anderson6141} for illustration due to the tunable system parameters,
e.g. double, triple and quadruple quantum dots (DQDs, TQDs and QQDs) connected to two biased reservoirs. The intrinsic features of the MBT and doublon are studied and their universality is addressed.

{\it The MBT.}---The CT is only related to the reduced time-dependent single-particle density matrix, $\rho(t)$. Thus the CT current is not sensitive to $U$. By contrast, the MBT engages at least reduced two-particle density matrix, $\pi(t)$. The two conjugated processes of the MBT are represented by $\pi_p(t)={\rm tr}[n_{i\bar{s}}c_{is}^{+}a_{js}\rho(t)]$ and $\pi_d(t)={\rm tr}[b_{js}^{+}c_{is}n_{i\bar{s}}\rho(t)]$ describing the dynamics of doublon production and decay respectively (\Fig{fig1}(a)), where $c_{is}$ ($c^{+}_{is}$),  with $s=\up,\down$,
denotes the annihilation (creation) operator matrix of an electron in the specified spin state in the $i$th dot; $n_{i\bar{s}}=c^{+}_{\bar{s}}c_{\bar{s}}$ is the number operator matrix with opposite spin of $s$; $a_{js}$ and $b_{js}$ denote dot or bath operator matrixes which can be the same or different\cite{Li12266403}. If spin flips during the doublon decay, the Kondo spin screening occurs  \cite{Kon6437,Ng881768}. In single QD, the Kondo resonance dominates the transport in the impurity magnetic moment regime, overwhelming the MBT by competition.

\begin{figure} 
\includegraphics[width=0.5\textwidth]{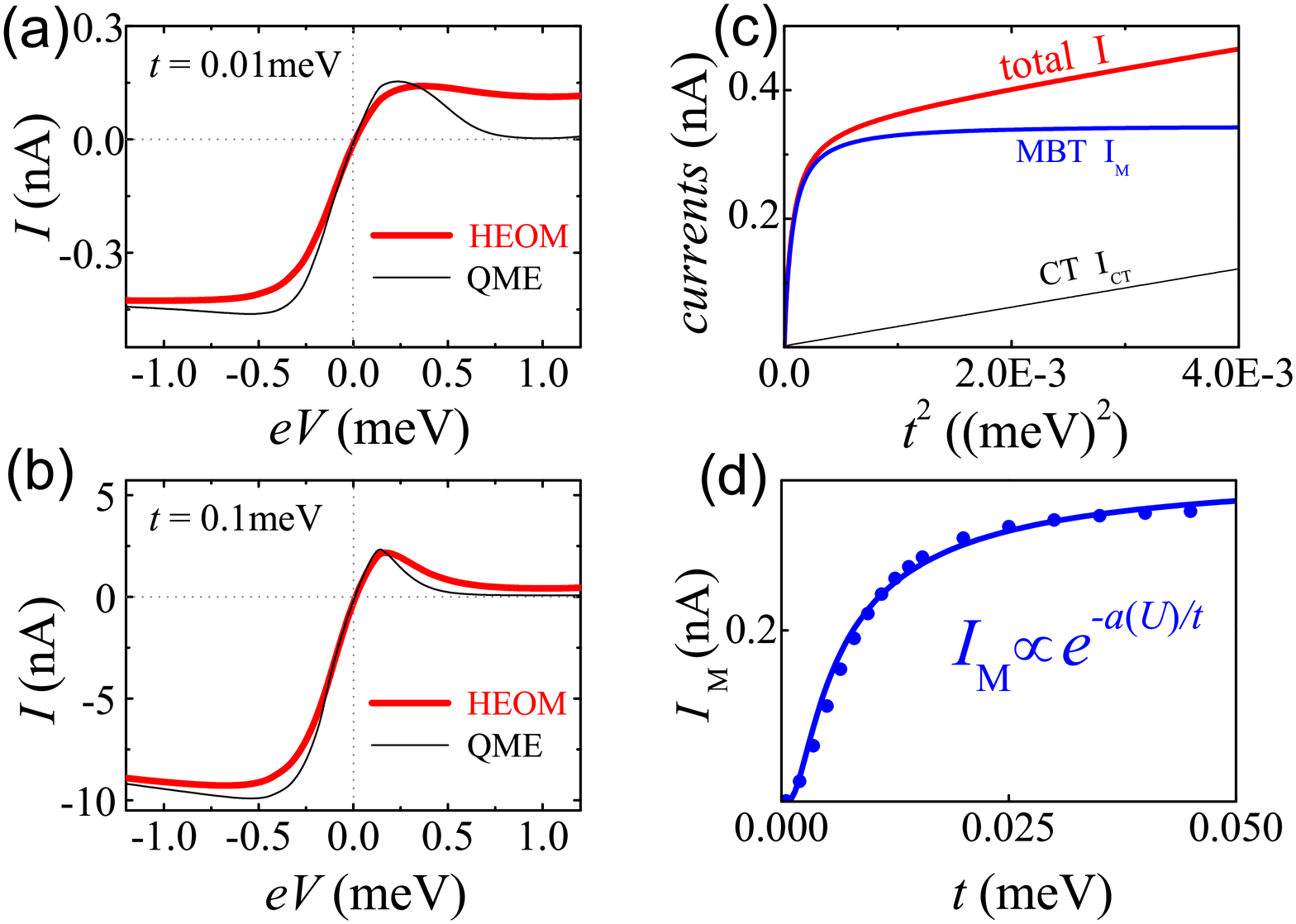}
\caption{(Color online). The $I-V$ curves are calculated at $t=0.01$ meV in (a), and $t=0.1$ meV in (b), for the DQDs, and $U=2.0\,{\rm meV}$ fixed. (c) The $I$, $I_{\M}$, and $ I_{\CT}$ and varying with $t^2$ are shown at $eV=U$. (d) The dependence of $I_{\M}$ on $t$ is given. The  dots are numerical data and the solid line is the plot of the function $I_{\M}\propto\exp[-\alpha(U)/t]$ with $\alpha(U)$ being a $U$-dependent parameter.}
\label{fig2}
\end{figure}

However, the MBT may make leading contribution in multi-QD systems. First of all, we introduce Pauli spin blockade (PSB) in DQDs. The PSB results from the Pauli's exclusion principle which prevents the spin triplet from transporting out of the DQDs through the transition to a singly or triply occupied state, and is characterized by the electron density accumulation as well as the current suppression in one bias direction \cite{Ono021313,Joh05165308,Liu08073310}. The reason of choosing the PSB regime is that the MBT dominates the transport leading to a finite current even when the CT is blocked. In fact, the current can be large enough to degrade the PSB effect, which can be a measurable symbol of the MBT (~\Fig{fig2}(a)).

To make the discussion concrete, we numerically calculate the nonequilibrium spectral functions $A_{is}(\omega)$ in the PSB regime, as shown in \Fig{fig1}(b) and (c).
The MBT  involves four events occurring almost simultaneously. i) A $\down$-spin electron tunnels from the left reservoir
to QD1; ii) The excess $\down$-spin tunnels from the QD1 to QD2; iii) The $\down$-spin electron in the QD2 excites to the doubly occupied level (doublon production); and iv) The doubly occupied $\down$-spin electron in the QD2 tunnels to the right reservoir (doublon decay). A key point is that the MBT strongly depends on $U$ in both doublon production and decay processes.
In comparison, the CT shown in \Fig{fig1}(c) can be well described by the second-order time-nonlocal quantum master equation (QME) \cite{Fra06205333}. Different to the MBT, the tunneling from the QD1 to the right reservoir is now via the singly occupied level of the QD2, without doublon production nor decay process. This leads to insensitivity of the CT to $U$, which is an essential difference of the CT to the MBT.

{\it Model and method.}---The Anderson multi--impurity model is adopted for $N$--QD ($N=$2, 3 or 4) systems. The total Hamiltonian reads $H_{\T}=H_{\s}+H_{\B}+H_{\SB}$.
The Hamiltonians of reservoirs   are
$H_{\B}=\sum_{\alpha ks}
(\varepsilon_{\alpha ks}+\mu_{\alpha})\hat{c}^\dg_{\alpha ks}\hat{c}_{\alpha ks}$, $\alpha=L,R$,
under the bias  $V=(\mu_{\rm L}-\mu_{\rm R})/e$, where $\hat c_{\alpha ks}$ ($\hat c^{\dg}_{\alpha ks}$)
denotes the annihilation (creation) operator of an electron in the
specified spin state in the $\alpha$-reservoir with wave vector $k$.
We set $E_F=\mu^{\rm eq}_{\rm L}=\mu^{\rm eq}_{\rm R}=0$ at equilibrium
and $\mu_{\rm L}/e=-\mu_{\rm R}/e=V/2$.
The system-reservoir coupling is
$H_{\SB}=\sum t_{\alpha kis}\hat c^{\dg}_{is}\hat c_{\alpha ks}+{\rm h.c.}$. The hybridization function is assumed to be a Lorentzian form
$J_{\alpha is}(\w)=\pi\sum_{k} t_{\alpha kis}t^{\ast}_{\alpha kis}
\delta(\omega-\varepsilon_{\alpha ks})
= \frac{\Gamma W^2}{\omega^2+W^2}$,
with $W=4\,{\rm meV}$ fixed in this letter.
The QD-reservoir coupling strength
$\Gamma$ will be specified below.
The Hamiltonian for central $N$-QD is
\[
 H_{\s}=\sum_{i=1,s}^{N} \epsilon_{is}\hat n_{is} + U\!\sum_{i} \hat n_{i\up} \hat n_{i\down}
      +t\!\sum_{<ij>,s}\!(\hat{c}^\dg_{is}\hat{c}_{js}+{\rm h.c.}).
\]
For DQD case, we lift the spin degeneracy of the QD1. This
can be achieved with
a local spin--splitting micromagnet \cite{Bru11146801},
resulting in $\mu_{\rm R}<\epsilon_{1\down}<\mu_{\rm L} < \epsilon_{1\up}$
for QD1.
The spin degeneracy of the QD1 will be restored later (see Supplemental Materials \cite{Hou16Supl}).
The spin degeneracy of the QD2 remains,
with $\epsilon_{2\up}=\epsilon_{2\down}$
and $\epsilon_{2s}<\mu_{\rm R}<\epsilon_{2s}+U<\mu_{\rm L}$.
The single--occupied levels, $\epsilon_{1s}$ and $\epsilon_{2s}$, are tuned via appropriate  local gate voltages to achieve the optimal PSB stability; i.e.,\ $\epsilon_{1\down}=\epsilon_{2s}+U=0$  \cite{Ono021313,Joh05165308,Liu08073310}. To study the long-range MBT in the TQD and QQD, a tilt energy $E_i$ is added to each dot as in ultracold atom experiments \cite{Meinert141259}, i.e., $H_{\s}=H_{\s}+\sum_{i,s} E_{i}\hat n_{is} $. For the tilted TQDs (see the insert of \Fig{fig4}(a)), $E_1=-E_3=E,\; E_2=0$; and for the tilted QQDs (see the insert of \Fig{fig4}(c)), $E_1=-E_4=3E/2,\;E_2=-E_3=E/2$, where $E$ is the nearest--neighbor tilt energy.

The multi-impurity Anderson model is solved using
the HEOM approach \cite{Jin08234703,Li12266403}.
This approach supports accurate and efficient evaluations of various steady--state and transient properties
\cite{Jin08234703,Wei2011arxiv,Li12266403,Hou14045141,Che15033009} (for more details, please refer to Ref.~[\onlinecite{YeWIREs}]  and references therein).  It has been demonstrated that the HEOM approach achieves the same level of accuracy as the latest high--level numerical renormalization group and quantum Monte Carlo approaches
for the prediction of various dynamical properties at equilibrium and nonequilibrium \cite{Li12266403}.

{\it MBT in DQDs.}---The current--voltage characteristics at $t=0.01$ and $0.1$ meV are shown in \Fig{fig2}(a) and (b), respectively. For comparison,
the QME results essentially from  the perturbative treatment are shown.
Let us examine closely the forward current, $I(V>0$).
In both $t=0.01$ meV and $t=0.1$ meV cases, the QME currents show negative differential conductances, due to the transition from the resonant to off--resonant tunneling \cite{Liu05161305,Wei07487}. At large $V$, the QME currents tend to near-zero constant value resulting from the CT, with a ratio of $I_{\rm peak}/I_{\rm const} \approx 50$. For precise HEOM currents (including the nonperturbative MBT currents), things are different. In $t=0.01$ meV case, the MBT current is increased and the negative differential conductance almost
vanishes. Although in the $t=0.1$ meV case, both the QME and HEOM currents display similar negative differential conductance, it should be pointed out that the HEOM current in \Fig{fig2}(b)  gives us $I_{\rm peak}/I_{\rm const}\approx 4$, {\it which matches the experimental data well} \cite{Liu05161305}. Although some other external mechanisms concerning the leakage current in the PSB regime have also been proposed in literatures \cite{Koppens051346}, here we provide a more intrinsic and universal one.

Figure \ref{fig2}(c) depicts the total current $I$,
and its two compositions,  $I_{\M}$ (for the MBT) and $I_{\CT}$ (for the CT),
as functions of $t^2$ at $U=2.0$ meV.
$I$ is the converged HEOM result, with $I_\CT$ for the time-nonlocal QME,
and $I_{\M}=I-I_{\CT}$. In contrast to $I_{\CT}\propto t^2$ resulting from the second-order perturbation theory,
$I_{\M}$ exhibits a strong nonlinearity before saturation.
By fitting the numerical data of $I_{M}$ in \Fig{fig2}(d),
we find $I_{\M}\propto \exp[-\alpha(U)/t]$, where $\alpha(U)$ is a parameter approximately proportional to $U$. It means that the MBT current has a similar dependence on $U/t$ to the decay rate of doublons observed in Ref.~[\onlinecite{Strohmaier10080401}], which  originates from the many-body character of the doublon dynamics.
It is more important that these features are rather general, beyond the case of $eV=U$ in \Fig{fig2}(c) and (d).

\begin{figure} 
\includegraphics[width=0.5\textwidth]{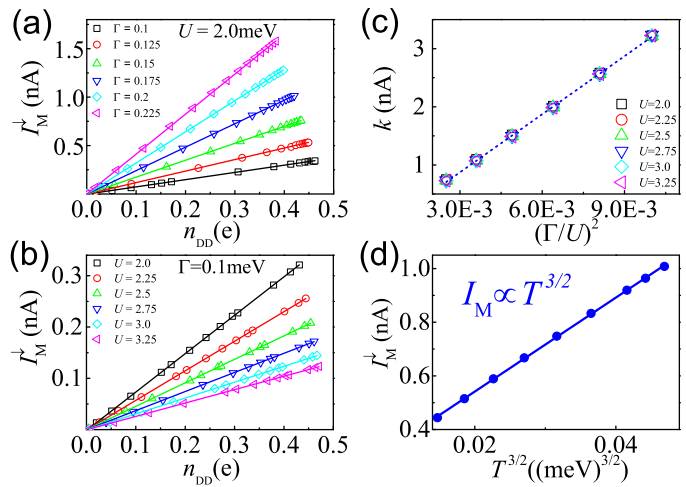}
\caption{(Color online).  The curves of $I^{\downarrow}_{\M}$ as a function of the self-consistent
 $n_{\DD}$ are shown, for different $\Gamma$  in (a), and different $U$ in (b). The energy unit in all figures is meV. In (a), $eV=U=2.0$ meV. In (b), $eV=U$  and $\Gamma=0.1$ meV. The scattered symbols are located by numerical data and the straight lines are plots of the function $I^{\downarrow}_{\M}=kn_{\DD}=k(1/2-n_{2\uparrow})$ with different slopes. The $k$ vs. $(\Gamma/U)^2$, and $I_{\M}=I^{\downarrow}_{\M}$ vs. $T^{3/2}$ are shown in (c) and (d) respectively. The symbols are from calculated data and the linear lines are fitting.
}
\label{fig3}
\end{figure}

For spin-nondegenerate-QD1 case, the tunneling current is basically carried by down spins, i.e.,
$I\simeq I^{\down}$, with $I^{\up}\simeq 0$. As shown in  \Fig{fig1}, $I^{\downarrow}_{\M}$ should be closely related to the number of doublons generated in QD2 and decayed into the right reservoir. We call this part of doublons as `dynamical doublons'  and denote their number as $n_{\DD}$.  In the DQDs, $n_{\DD}(t)= 1/2-n_{2\up}(t)$, where $n_{2\up}$ is the $\up$-spin occupation number in QD2. For the uncoupled limit $n_{2\up}(t=0)=1/2$ leading to $n_{\DD}(t=0)=0$. $I^{\down}_{\M}$ is evaluated as a function of $n_{\DD}$
for different $\Gamma$ and $U$ in \Fig{fig3}(a) and (b), respectively.  Remarkably, $I^{\downarrow}_{\M}$ keeps a fundamental linear function of $n_{\DD}$, i.e.
\be  \label{eq1}
  I^{\downarrow}_{\M}=kn_{\DD}=k\left(\frac{1}{2}-n_{2\uparrow}\right),
\ee
in all regimes of $t$, across a broad range of $U$, $\Gamma$, and temperature $T$. For the parameters chosen in \Fig{fig2}(c), $k=0.75$ nA.
At $t>0.05$ meV, $n_{2\uparrow}=0.05$ and $n_{\DD}=0.45$,  thus $I^{\downarrow}_{\M}=0.33$ nA, which is completely consistent with the saturation current shown in \Fig{fig2}(c).  $I_{\M}\propto n_{\DD}$ proves that the dynamical doublons are the only carriers of $I_{\M}$, namely, the MBT precisely describes the dynamics of doublons, as we argued above.

The ideal linear plots in \Fig{fig3}(c) shows further scaling laws of the slope $k$ on the ratio of $\Gamma/U$, $k\propto(\Gamma/U)^2$. In \Fig{fig3}(d), we prove the temperature dependence of the MBT current as,  $I_{\M}(=I^{\downarrow}_{\M})\propto T^{3/2}$, which is distinctly different from the $T^2$-dependence of $I_{\CT}$  \cite{Naz09}.  This unconventional $T$ dependence, originally attributed to the intrinsic many-body character of the MBT, can be conveniently verified by experiments.

It is necessary to extend the study to the spin-degenerate
case in which the local Zeeman splitting in QD1 is removed.
Numerical calculations show that the MBT current is not sensitive to this spin splitting
and behaves in a similar manner to the spin--nondegenerate case;
see the Figs.\,S1-S3 of Ref. \cite{Hou16Supl}.

\begin{figure} 
\includegraphics[width=0.5\textwidth]{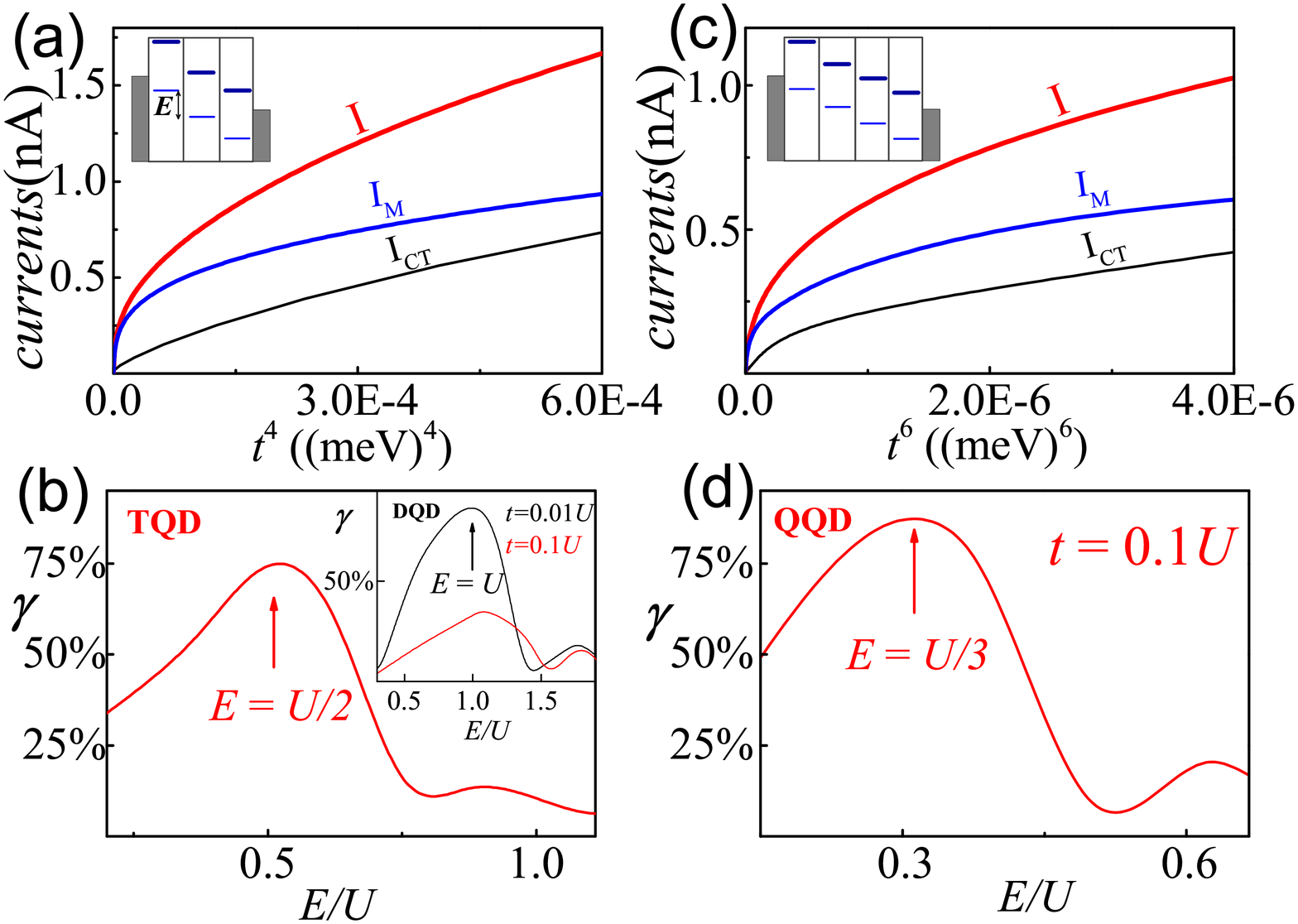}
\caption{(Color online). (a) The dependence of $I$, $I_{\M}$ and $ I_{\CT}$ on $t^4$ at $eV=1.5$ meV for the TQDs with tilting energy $E=U/2$.  (b) The dependence of $\gamma\equiv (I_{\M}/I)\times100\%$ on the ratio of $E/U$ in the TQDs at $t=0.1U$ is displayed. The insert shows $\gamma$ as a function of $E/U$ in the DQDs at $t=0.01U$ and $0.1U$. (c) The  $I$, $I_{\M}$ and $ I_{\CT}$  vary with $t^6$ at $eV=1.5$ meV for the QQDs with $E=U/3$. (d) The $\gamma$ as a function of $E/U$ is depicted for the QQDs at $t=0.1U$. The other parameters are $\epsilon_{i,s}=-U/2=-1.0$ meV (TQD: $i=1-3$; QQD: $i=1-4$), $\Gamma=0.1$ meV and $T=0.1$ meV.}\label{fig4}
\end{figure}

{\it Long-range MBT.}---Now, we are on the position to demonstrate the coherent long-range MBT in larger systems than DQDs.  In recent experiments of the Bose-Hubbard chain at $t\ll U$, long-range tunneling over up to five sites were observed as resonances in the number of doublons when the nearest-neighbour tilt energy ($E$) is tuned to integer fractions of $U$ \cite{Meinert141259}. Let us prove that the observations arise from the long-range MBT by numerical results in tilted TQDs and QQDs (see inserts in \Fig{fig4}(a) and (c)). \Fig{fig4}(a) depicts the dependence of  $I$, $I_{\M}$ and $ I_{\CT}$  on $t^4$ at $eV=1.5$ meV for the spin degenerate TQD with  $E=U/2$. Not surprisingly, all of them behave in a similar manner as those in DQD. Since the number of tunneling barriers doubles now, we have $I_{\CT}\propto t^4$.

In tilted QDs, $I_{\CT}$ can not be blocked as that in the PSB region, thus it is mixed with $I_{\M}$ in observed $I$. In order to separate out the direct contribution of the MBT, we define $\gamma\equiv (I_{\M}/I)\times100\%$. \Fig{fig4}(b) shows the dependence of $\gamma$ on the ratio of $E/U$ for the TQDs. The case of titled DQDs is shown in the insert for the purpose of comparison. For the TQDs, a resonant peak at $E=U/2$ is clearly seen; while for the DQDs, that peak shifts to $E=U$, with the peak position unchanged at different $t$. The maximum value of $\gamma$ up to $75\%$ in TQDs further proves that the resonance at $E=U/2$ results from the long-range MBT rather than the CT which in fact has been excluded already by the strong $U$-dependence of the resonance. Both the peak structure and the peak position for the DQDs and TQDs shown in \Fig{fig4}(b) are well consistent with experimental observations for Cs atoms in optical lattice in Ref.~[\onlinecite{Meinert141259}], which suggests that our theory shares the same physics of long-range tunneling of doublons in experiments about ultracold atoms although their objects of study are quite different.

Our theoretical results for titled QQDs  further confirm the above arguments. In \Fig{fig4}(c),  $I_{\CT}$ is approximately proportional to $t^6$ at $E=U/3$. In \Fig{fig4}(d), a similar peak structure is clearly seen with the resonant peak located at $E=U/3$, which is exactly the position of resonant doublons tunneling through four sites of titled Hubbard chain, as observed in expriments \cite{Meinert141259}. We thus conclude that the MBT can support the coherent long-range tunneling of doublons in general systems, which is of both fundamental and practical importance. For example, the MBT can provide a feasible way to manipulate distant quantum gates or qubits in one step in solid-state quantum computing. This process would enhance the operating efficiency and fault-tolerant capability, compared to the nearest-neighbor control in exchange-based quantum gates \cite{Los98120}.

{\it Summary.}---In summary, we have theoretically discussed the many-body tunneling (MBT), by nonperturbatively solving the Anderson multi-impurity model, and identified it a fundamental tunneling process that  involves the dynamics of doublons. Proportional to the numbers of dynamical doublons, the MBT is shown to dominate the off-resonant transport in strongly correlated systems. A $T^{3/2}$-dependence of the MBT current is uncovered and can be identified as a fingerprint of MBT in experiments. It is also proved that the MBT can support the coherent long-range tunneling of doublons. Our theoretical results are well consistent with recent experiments on the dynamics of doublons. As a fundamental physical process, the MBT is expected to play important roles in more general systems beyond what we have discussed here.

The support from the NSFC (Grant No.~11374363),
Strategic Priority Research Program (B) of the CAS (No.~XDB01020000)
and Research Funds of Renmin University of China (No.~11XNJ026)
is gratefully appreciated. ZGZ is supported
by the Hundred Talents Program of CAS.
Computational resources have been provided by the
Physical Laboratory of High Performance Computing at Renmin University of China.

\end{document}